\documentclass{svproc}
\usepackage{url}

\usepackage{amsmath,amssymb}

\usepackage{changepage}

\usepackage[utf8x]{inputenc}

\usepackage{textcomp,marvosym}

\usepackage{cite}

\usepackage{nameref}

\usepackage[table]{xcolor}

\usepackage{setspace} 

\usepackage{lastpage,fancyhdr,graphicx}
\usepackage{subfig}

\begin{document}
\mainmatter              
\title{From mean-field to complex topologies: network effects on the algorithmic bias model}
\titlerunning{network effects on the algorithmic bias model}  
%
\author{Valentina Pansanella\inst{1,3} Giulio Rossetti\inst{3} Letizia Milli\inst{2,3}
}
\authorrunning{Valentina Pansanella et al.} 
%
%
\institute{Scuola Normale Superiore, Pisa, Italy\\
\email{valentina.pansanella@sns.it},\\ \and Department of Computer Science, University of Pisa, Italy \\ \email{milli@di.unipi.it}
\and KDD Lab ISTI-CNR, Pisa, Italy\\ \email{\{name.surname\}@isti.cnr.it}}

\maketitle              

\begin{abstract}
Nowadays, we live in a society where people often form their opinion by accessing and discussing contents shared on social networking websites.
While these platforms have fostered information access and diffusion, they represent optimal environments for the proliferation of polluted contents, which is argued to be one of the co-causes of polarization/radicalization. 
Moreover, recommendation algorithms - intended to enhance platform usage - are likely to augment such phenomena, generating the so called \emph{Algorithmic Bias}. In this work, we study the impact that different network topologies have on the formation and evolution of opinion in the context of a recent opinion dynamic model which includes bounded confidence and algorithmic bias. Mean-field, scale-free and random topologies, as well as networks generated by the Lancichinetti-Fortunato-Radicchi benchmark, are compared in terms of opinion fragmentation/polarization and time to convergence. 
\end{abstract}
\section{Introduction}

One of the most analyzed phenomena on online social networks is the tendency to observe political polarization, e.g., online discourse's tendency to divide users into opposite political factions not aiming at reaching any form of synthesis. Polarization not only emerges in political debates, but it also often characterizes a variety of controversial topics, and, in some cases, it may affect policymaking and society. 

The tendency to observe political polarization, e.g., online discourse's tendency to divide users into opposite political factions has captured great interest recently \cite{Campbell20181}, affecting also polarization on a wide variety of controversial topics and eventually policymaking and society.

Among the various causes of this phenomenon, there is considered to be the rise of social media, in particular because of the presence of personalization/recommendation algorithms \cite{SPGK19
}. 
Even if their initial intent was to maximize platform usage and users' engagement, these algorithms end up working as a reinforcement bias for online users' opinions, neglecting them access/confront with narratives different from their own. 

The field of opinion dynamics aims at understanding the opinion formation, evolution, and eventually stabilization in groups of interacting agents. Typically, such models consider a finite number of interacting people (
\emph{agents}); each agent has its own opinion, discrete or continuous, which can vary over time, 
according to rules to explain the change due to interactions with other agents.

The idea of studying human behavior in the same way that physics is studied dates back to the $18^{th}$ century with the work of the philosopher David Hume\cite{DY1739}. This science is referred to as \lq\lq sociophysic\rq\rq \cite{Schweitzer_2018}, and it draws on the idea that general laws are describing human societies and human behavior; therefore, the statistical analysis can be performed on human qualities to explain human behavior.
The advent of Big Data stimulated a renewed interest in this field, first regarding financial markets and then exploiting data from the internet and online social networks. 

The last decade witnessed the introduction of a wide variety of models that start from the milestones \cite{Fre56, DeG74, Fri86, DW00} and extend them, considering different sociological theories that try to 
incorporate the new characteristics of today's society that influence group dynamics \cite{S_rbu_2016}. 
With the development of online social networks, more and more works tried to include their specific characteristics in the opinion dynamics model to get closer to reality.
Filtering algorithms \cite{SPGK19
}, fake news, underlying complex networks with specific topologies and communities, 
are considered essential elements to implement a good opinion dynamics model, and researchers are surpassing the idea of complete network structures and synchronous interactions. 

In this paper, we investigate the expected effects of algorithmic bias in a networked population.
Moving from the results discussed in \cite{SPGK19} where a mean-field context is assumed 
(e.g., all individuals can interact among them without any social restrictions), 
we aim to study the effect that different network topologies have on opinion formation and evolution when in the presence of a filtering algorithm.

To such extent, and to allow results reproducibility, we focus our analysis on well-known network models, namely Erdős–Rényi\cite{ER59}, Barabási–Albert\cite{BA99} and Lancichinetti–Fortunato–Radicchi benchmark \cite{LFR08} graphs (henceforth referred to as LFR graphs).

Adopting such controlled environments, used to simulate the social structure among a population of interacting individuals, we analyze the behaviors of the Algorithmic Bias model\cite{SPGK19} (e.g., an extension of the Deffuant-Weisbuch\cite{DW00} opinion dynamics model) and discuss the role of graph properties on the observed simulations results.

\smallskip
The paper is organized as follows. 
In section \ref{sec:modelingcomplex} we introduce the algorithmic bias model, and we describe our experimental workflow. 
Section \ref{sec:results} discusses the main finding of our simulations, finally section \ref{sec:conlcusion} concludes the paper while opening to future investigations.

\section{Algorithmic bias: from mean-field to complex topologies}
\label{sec:modelingcomplex}
Considering such a complex scenario, the present work aims to deepen the AB model's behavior analysis and test it on different network topologies. 

Nowadays, online social networks have become the primary source of information and an excellent platform for 
opinion exchanges. 
However, the flow of content that each user sees is organized by algorithms that are built to maximize platform usage: from this, it comes to the hypothesis that there is an algorithmic bias (also called algorithmic segregation) since these contents are selected based on users' precedent actions on the platform or the web, 
reinforcing the human tendency to interact with content confirming their beliefs (confirmation bias). 

To introduce in the study of opinion dynamics the idea of a recommender system generating an algorithmic bias, we started from a recent extension of the well-known Deffuant-Weisbuch model (DW-model henceforth), proposed in \cite{SPGK19} (Algorithmic Bias model or AB model, henceforth).

In the Algorithmic Bias model, we have a population of $N$ agents, where each agents $i$ has a continuous opinions $x_{i} \in [0,1]$. 
At every discrete time step the model randomly select a pairwise $(i, j)$, and, if their opinion distance is lower than a threshold $\epsilon$, $|x_{i} - x_{j}| \leq \epsilon$, then the two agents change their opinion according to the following rule:
\begin{equation}
\begin{aligned} 
    \label{eq:updateDW}
    x_{i}(t+1) &=  x_{i} + \mu(x_{j}-x_{i}) \\ 
    x_{j}(t+1) &=  x_{j} + \mu(x_{i}-x_{j}).
\end{aligned}   
\end{equation}

The parameter $\epsilon \in [0,1]$ models the population's confidence bound, and it is assumed to be constant among all the agents. 
The parameter $\mu \in (0, 0.5]$ is a convergence parameter, modeling the strength of the influence the two individuals have on each other or, in other words, how much they change their opinion after the interaction. 
The numerical simulations of this model show that the qualitative dynamic is mainly dependent on $\epsilon$: as $\epsilon$ grows, the number of final opinion clusters decreases. 
As for $\mu$ and $N$, these parameters tend to influence only the time to convergence and the final opinion distribution width. 

The AB model introduces another parameter to model the algorithmic bias: $\gamma \geq 0$. This parameter represents the filtering power of a generic recommendation algorithm: if it is close to $0$, the agent has the same probability of interacting with all its peers. As $\gamma$ grows, so does the probability of interacting with agents holding similar opinions, while the probability of interacting with those who hold distant opinions decreases.
Therefore, this extended model modifies the rule to choose the interacting pair $(i, j)$ to simulate a filtering algorithm's presence. An agent $i$ is randomly picked from the population, while $j$ is chosen from $i$'s peers according to the following rule:
\begin{eqnarray}             
    \label{eq:algbias}
    p_{i}(j)=\frac{d_{ij}^{-\gamma}}{\sum_{k \ne i}d_{ik}^{-\gamma}}
\end{eqnarray}
where $d_{ij} = |x_{i}-x_{j}|$ is the opinion distance between agents $i$ and $j$, so that for $\gamma = 0$ the model goes back to the DW-model.
When two agents interact, their opinions change if and only if the distance between their opinions is less than the parameter $\epsilon$, i.e. \(|x_{i}-x_{j}| \leq \epsilon \), according to Eq.~\ref{eq:updateDW}.
In~\cite{SPGK19} the discussed AB model results focus only on the mean-field scenario, i.e., the authors assume a complete graph as the underlying social structure.

On online social networks, each agent will likely interact only with whom they follow or with their friends. 

Our goal is to verify if/how networks' structure exacerbates the polarization and fragmentation generated by the selection bias's presence. 
We want to verify if moving from a complete network with $L_{max}$ links to a network with $L<<L_{max}$ links and a predetermined topology influences the final simulation state, making it harder for the population to reach a consensus or ultimately preventing it.

To such an extent, we first replicate the work of \cite{SPGK19} on a complete network (to define a reliable baseline), and after that, we show the AB evolution on two different graph models: random\cite{ER59} and a scale-free\cite{BA99} network. 
We extend the experimentation carried out in \cite{SPGK19} because the mean-field assumption neglect key properties of real networks, such as the sparsity and the small-world phenomenon (captured by the ER model\cite{ER59}) and the scale-free degree distribution (captured by the BA model\cite{BA99}).

Moreover, to understand the impact of realistic mesoscale network topology (e.g., presence of communities) on opinion evolution, we also test the AB model against a network generated through the LFR benchmark \cite{LFR08}. 
Our research aims to investigate whether, in realistic environments, opinions remain trapped inside communities or not and which are the effects of different topologies on the steady-state of the modeled dynamic process, e.g., whether they facilitate/counteract polarization/fragmentation or promote consensus. 

In all scenarios, we set the number of nodes $N=250$.
For the ER network, we fix the $p$ parameter (probability to form a link) to $0.1$ (thus imposing a \emph{supercritical} regime, as expected from a real-world network); we obtain a random network composed of a single giant component with an average degree of $24.94$.
In the BA network, we set the $k$ parameter (number of edges to attach from a new node to existing nodes) to $5$, thus creating a network with an average degree equal to $9.8$.

We generated nine different networks using the LFR benchmark ($N=250$). The parameters used for its construction have been set as follows: 
\begin{itemize}
    \item power-law exponent for the degree distribution, $\gamma = 3$; \item power-law exponent for the community size distribution, $\beta = 1.5$; \item fraction of intra-community edges incident to each node, \\
    $
        \mu_{LFR} \in \{ 0.1, 0.5, 0.9\}
    $; \item average degree of nodes, $< k >$ = 10; \item minimum community size $min_s = 50$, thus losing the power-law community size distribution and generating $4$ communities of similar sizes in the end.
\end{itemize}
The parameter $\mu_{LFR}$ controls the number of edges between communities, thus reflecting the network's amount of noise. Therefore, the network with $\mu_{LFR} = 0.1$ has better-defined communities than the one generated with $\mu_{LFR} = 0.9$.

\section{Experimental analysis and results}
\label{sec:results}
Like in \cite{SPGK19}, to avoid undefined operations in equation \ref{eq:algbias}, when $d_{ik} = 0$ we use a lower bound $d_{\epsilon} = 10^{-4}$. 
The simulations are designed to stop when the population reaches an equilibrium, i.e., the cluster configuration will not change anymore, even if the agents keep exchanging opinions. 
We also set an overall maximum number of iterations at $10^5$. 
To account for the model's stochastic nature, we compute the average results over $10$ independent executions for each configuration, where the initial opinion distribution is always drawn from a random uniform probability distribution in [0,1]. 
To better understand the differences in the final state concerning the different topologies considered, we study the model on all networks for different combinations of the parameters. 
We are interested in whether, parameters being equal, the different topology influences the final cluster configuration enhancing polarization and fragmentation, but also the dynamics of the process, by slowing down the convergence or reducing the density of the final opinion clusters. 
In the simulations, we tested the model on every possible combination of the parameters over the following values:
\begin{itemize}
    \item $\epsilon$ takes a value from 0.2 to 1.0 with the step of 0.1.
    \item $\gamma$ takes value from 0 to 2.0 with the step of 0.2; for $\gamma = 0$ the model becomes the DW-model.
    \item $\mu = 0.5$, so whenever two agents interact, if their opinions are close enough, they update to the pair's average opinion.
\end{itemize}

The AB model implementation used to carry out our experiments is the one provided by the NDlib \cite{rossetti2018ndlib} Python library\footnote{NDlib: \url{http://ndlib.rtfd.io}}.

For the simulations of the AB model on the LFR benchmark networks, instead, we tested the model over the following values:
\begin{itemize}
    \item $\epsilon \in \{0.2, 0.3\}$. We impose this choice because in the mean-field, for these values, the number of clusters grows with increasing gamma, and we obtain a situation of polarization and fragmentation:
    \item $\gamma \in \{0.0, 0.5, 1.0, 1.5, 2.0\}$;
    \item $\mu = 0.5$. With this value, whenever two agents interact, if their opinions are close enough, they update to the pair's average opinion.
\end{itemize}

\begin{figure}[h]
    \includegraphics[width=1.0\textwidth]{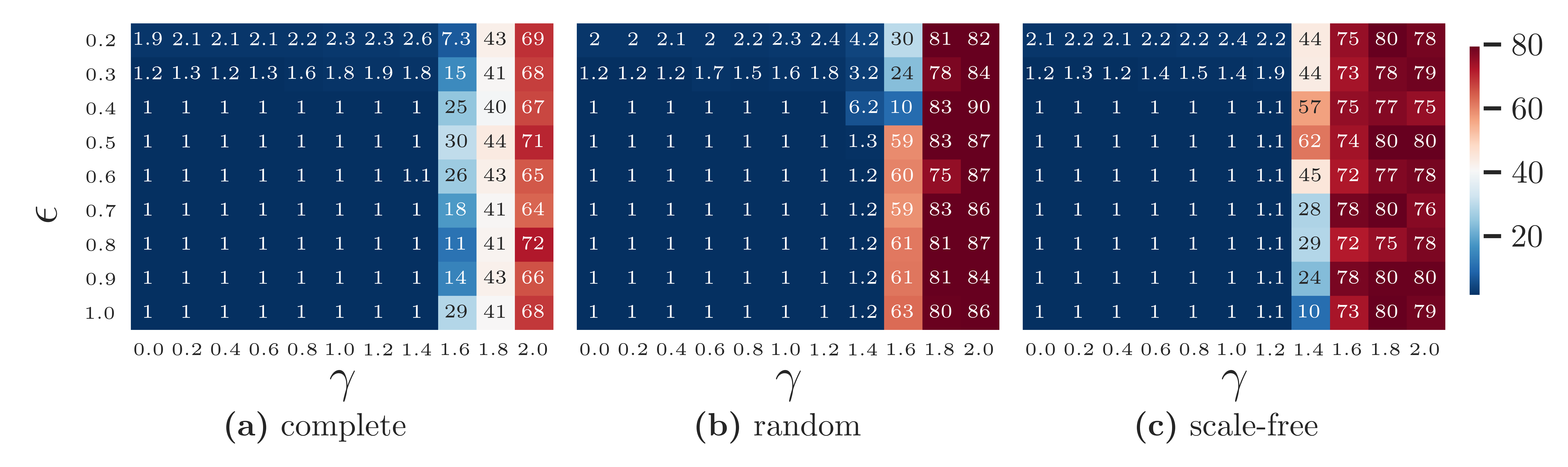}
    \caption{\textbf{Average number of clusters across topologies.} The figure displays the average number of clusters as a function of $\epsilon$ and $\gamma$, over $10$ runs. We show the results both for $\gamma$ from 0 to 2.0 with step of 0.2 and $\epsilon$ from 0.2 to 1.0 with step 0.1 respectively for a complete network (a) a random network (b) and a scale-free network (c)}
    \label{fig:nclusterheatmaps}
\end{figure}

\paragraph{Average number of clusters.}To analyze the results of the simulation we start by taking into account the number of final opinion clusters in the population, to understand the degree of fragmentation that the different combination of the parameters produce.  
This value indicates how many peaks there are in the final distribution of opinions and provides a first approximation of whether a consensus can be obtained or not.
To compute the effective number of clusters, accounting for the presence of major and minor ones, we use the cluster participation ratio, as in \cite{SPGK19}:
\begin{eqnarray}
    \label{eq:ncluster}
    C = \frac{(\sum_{i}{c_{i}})^{2}}{\sum_{i}{c_{i}^{2}}}
\end{eqnarray}
where $c_{i}$ is the dimension of the $i$th cluster, i.e., the fraction of population we can find in that cluster. In general, for $m$ clusters, the maximum value of the participation ratio is $m$ and is achieved when all clusters have the same size, while the minimum can be close to 1, if one cluster contains most of the population and a very small fraction is distributed among the other $m$ − 1.\\
From fig. \ref{fig:nclusterheatmaps} we can see that the behavior of the model across the different network topologies is very similar: the growth of the confidence bound $\epsilon$ allows the population that initially ended up as polarized to reach a full and perfect consensus, at least up to a certain value of the algorithmic bias $\gamma$. The experiments on the three different networks show how the population either converges to one or a few significant clusters or fragments over a wide range of opinions, when $\gamma$ is above a certain threshold: the final state shows tens of clusters populated by few agents that cannot merge in the time span allowed in these experiments. 

Even in the mean-field for $\gamma \geq 1.6$ the effect of the algorithmic bias is too strong to be mitigated by an increment in the bounded confidence parameter $\epsilon$. The total number of clusters grows with $\gamma$ from values around $10$ to values around $70$. 

However, the population only has $10^5$ iterations to reach convergence and in some cases the process reaches this bound without having reached equilibrium, like we will see later in this paragraph. In \cite{SPGK19}, instead, the maximum number of iterations was set to $10^7$, allowing the population to always reach a steady-state. While analysing what happens when time goes to infinite is important, it's also important to understand how the final status may change with a much shorter dynamic. The present results could mean that consensus - even if theoretically possible - may never be reached in a real setting where there is a finite amount of time to discuss a topic and the population may instead remain fragmented. 

Considering only $\gamma \leq 1.4$ we can see that up to this value, the results remain the same described in \cite{SPGK19}: for $\epsilon \geq 0.4$ a consensus is always reached, even if it tends to become less and less perfect, while for $\epsilon \le 0.4$ the number of clusters increases with the bias, which brings the population to a polarization of opinions even in situations where the DW-model \cite{DW00} would have produced a full consensus.

Introducing a different network topology, such as a random network or a scale-free network, however, produces a change in the behaviour for very strong biases. 
Such a result suggests that a sparser topological structure has a small impact on the observed results until the introduced bias is not strong.
However, as the algorithmic filtering grows stronger, the sparsity has a very severe impact, preventing consensus - even in cases where it was observed as a possible outcome in mean-field. 
Moreover while for $\gamma$ values below the fragmentation threshold the effective number of clusters is very similar across the three different network topologies, in the fragmented state we can see that in the scale-free case the number of clusters is higher - on average - than in the random case and both show overall higher values with respect to the complete networks. It is not clear how this different behaviour depends on the topology and how it depends on a different average degree and thus total number of links in the networks, but we can assume that the more the sparsity, the more it gets difficult for opinion clusters to merge when the bias limits very much the number of agents to interact with. 
 
\begin{figure}[ht]
\includegraphics[width=1.0\textwidth]{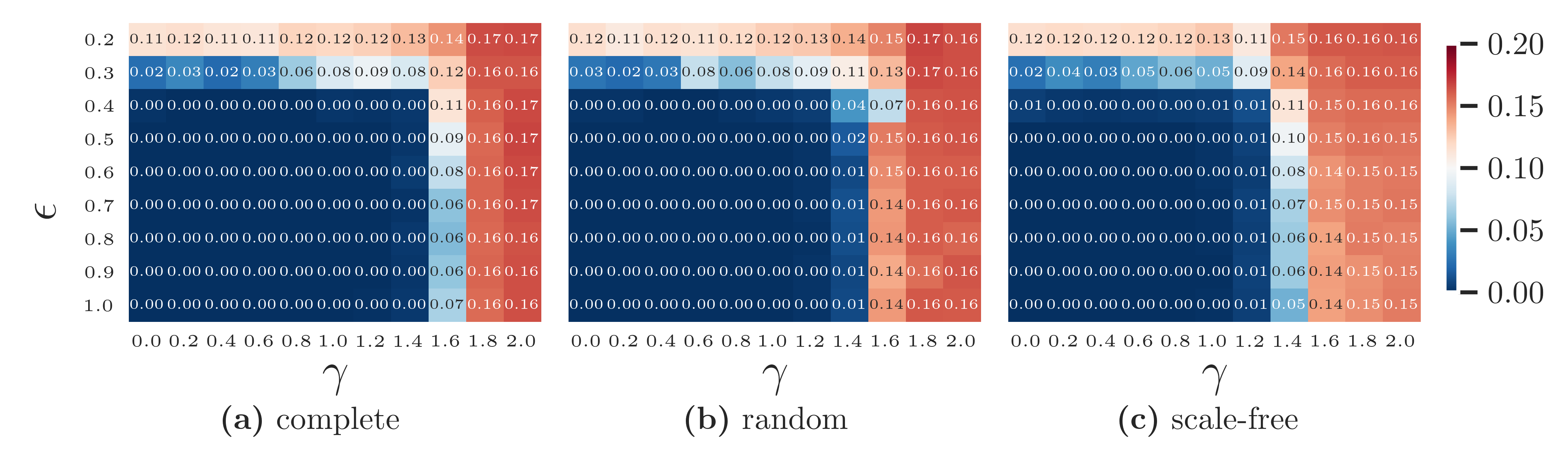}
\caption{\textbf{Average opinion distance across topologies.} The figure displays the average pairwise opinion distance as a function of $\epsilon$ and $\gamma$, over $10$ runs. We show the results both for $\gamma$ from 0 to 2.0 with step of 0.2 and $\epsilon$ from 0.2 to 1.0 with step 0.1 respectively for a complete network (a) a random network (b) and a scale-free network (c)}
\label{fig:pwdistheatmaps}
\end{figure}

\paragraph{Average pairwise distance. }To study the degree of polarization/fragmentation, we computed the average pairwise distance between the agents' opinions. 
Given an agent $i$ with opinion $x_{i}$ and an agent $j$ with opinion $x_{j}$ at the end of the diffusion process, the pairwise distance between the two agents is $d_{ij} = | x_{i} - x_{j} |$. 
The average pairwise distance in the final state can be computed as ${\frac{\sum_{i=0}^{N}{\frac{(\sum_{j=0}^{N}{d_{ij}})}{N}}}{N}}$.
In every network, the average opinion distance goes from a minimum value of $0.0$, when the population reaches a full consensus and every agent holds the same opinion, to a maximum value of $0.15-0.17$ when there are tens of opinion clusters in the population. 
We can observe that such a distance follows the same pattern identified in the number of clusters: it decreases as $\epsilon$ grows and grows with $\gamma$. 
However, an important thing to point out is that while the difference in the number of clusters can be very high, the opinion distance differences are not so high between a state with 3 final clusters and a state with 80 final clusters. 
Indeed, when the opinion distribution is very fragmented, the different clusters tend to get closer to each other. This holds for the three different networks considered in this work. 

There are also some cases in the complete network where the average pairwise distance decreases despite the number of clusters in the final state is higher. This result suggests that the final peaks in the opinion distribution are indeed all very close to each other and with a longer simulation a lower level of fragmentation could be reached. 

That considered, we can state that the average pairwise distance is suitable to highlight the transition from consensus to polarisation or a limited number of clusters. 
However, it is not suitable to characterize a growing fragmentation or an intermediate state where there are still many cluster growing closer to each other before merging. 

\paragraph{Time to convergence. }Finally, we consider the time to convergence.
The time to convergence is measured as the number of iterations (each constituted by $N$ pairwise interactions).
\begin{figure}[ht]
    \includegraphics[width=1.0\textwidth]{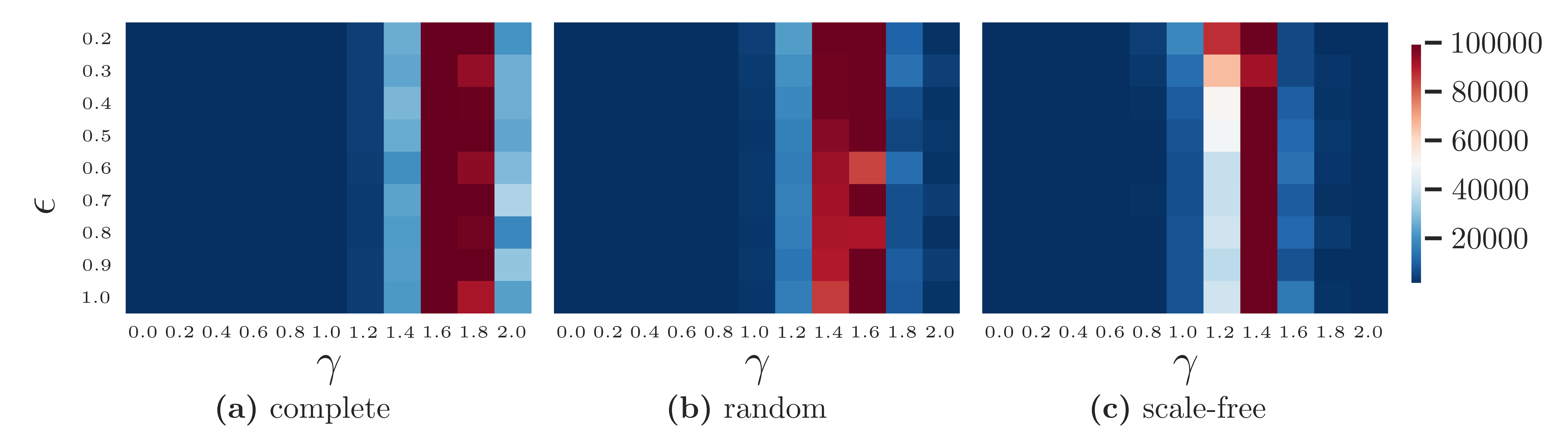} 
    \caption{\textbf{Average number of iterations to convergence across topologies.} The figure displays the average number of iterations to convergence as a function of $\epsilon$ and $\gamma$, over $10$ runs. We show the results both for $\gamma$ from 0 to 2.0 with step of 0.2 and $\epsilon$ from 0.2 to 1.0 with step 0.1 respectively for a complete network (a) a random network (b) and a scale-free network (c)}
    \label{fig:niterheatmaps}
\end{figure}

Fig. \ref{fig:niterheatmaps} compares the evolution of the time to convergence as a function of $\epsilon$ and $\gamma$. The three plots all show a similar behavior: the main impact on time to convergence (since $\mu$ and $N$ are fixed) is given by $\gamma$.  
In particular, for every value of the parameter $\epsilon$ in every network the convergence slows down until it reaches its peak for a certain value of the bias, then the time to convergence starts to decrease as the bias grows. 

\paragraph{Mesoscale structure. }
We saw from the previous analysis that changing the topology of the network, even with a low average degree and a scale-free structure - doesn't affect much the dynamics.
To understand how - instead - the addition of a mesoscale structure may affect the process of opinion diffusion we simulated three different scenarios over the previously described LFR networks and we analyzed the same measures as a function of $\gamma$, $\epsilon$ and also $\mu_{LFR}$. 
We fixed three different settings: 
\begin{itemize}
    \item[i)] the opinions are uniformly distributed across the whole population, like on complete, random, and scale-free cases previously analyzed;
    \item[ii)] a random mean opinion to each community is assigned, and then the opinions are normally distributed within the community with standard deviation equal 0.01;
    \item[iii)] the opinions are normally distributed with predefined means \\ $\in \{0.25, 0.5, 0.75, 1.0\}$ and a standard deviation equal $0.01$. 
\end{itemize}

\begin{figure}[ht]
    \centering
    {\includegraphics[width=0.45\textwidth]{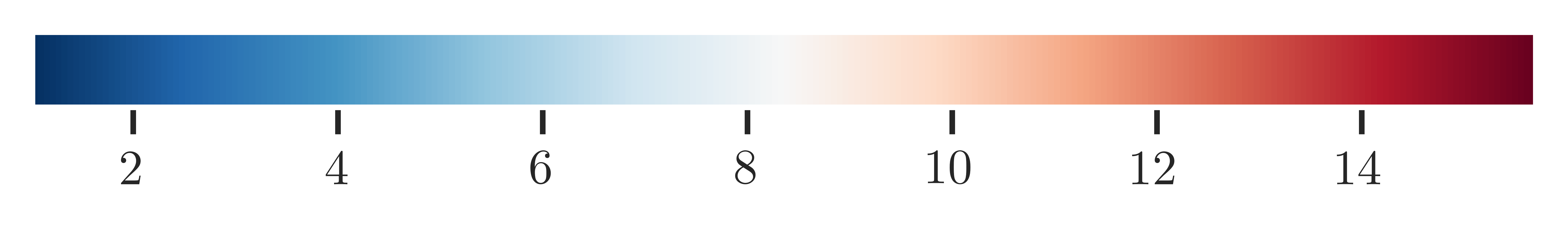}}\\

\subfloat[\centering Uniform distribution $\epsilon=0.2$]
{ \includegraphics[width=0.3\textwidth]{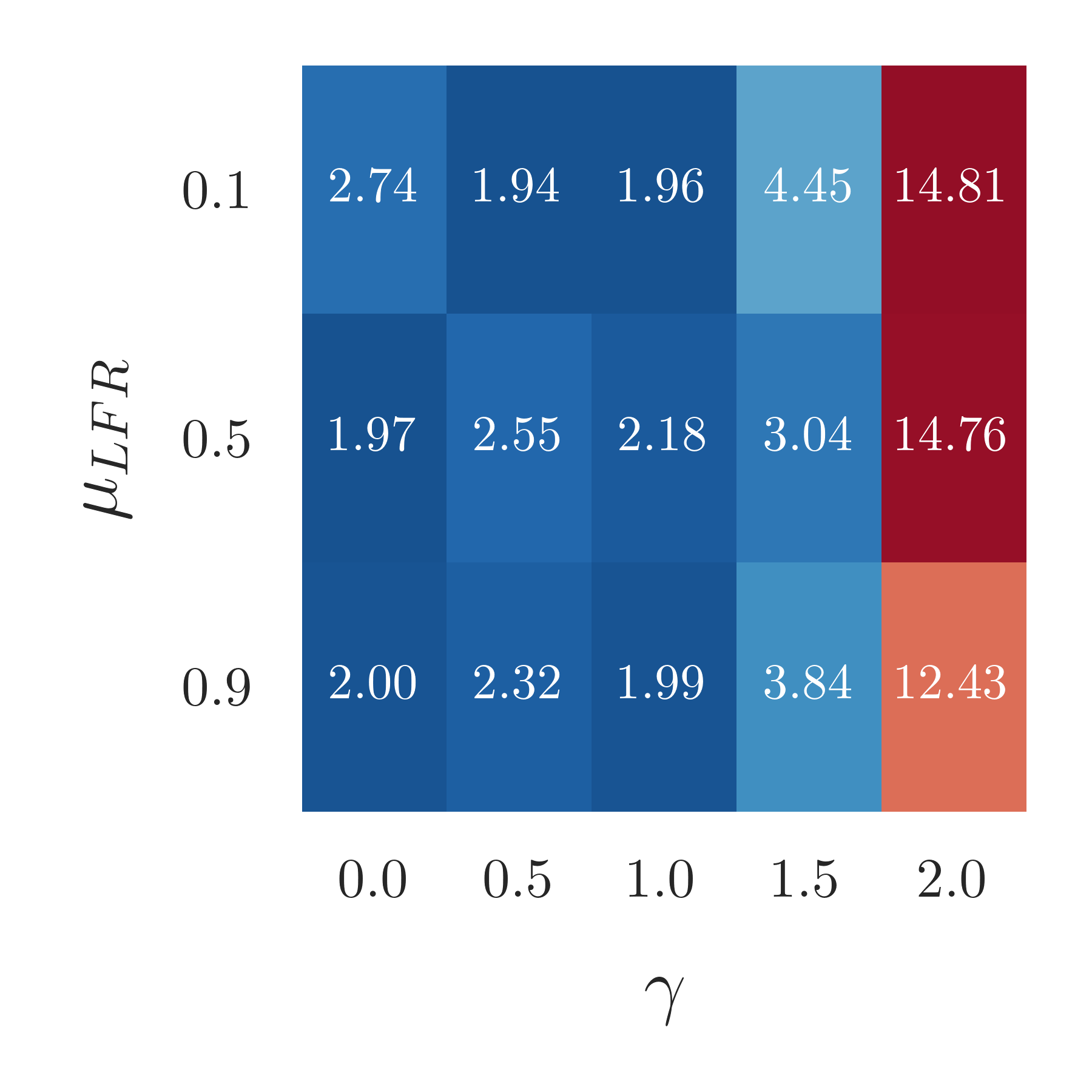}}     \hfill
\subfloat[\centering Random means $\epsilon=0.2$]{ \includegraphics[width=0.3\textwidth]{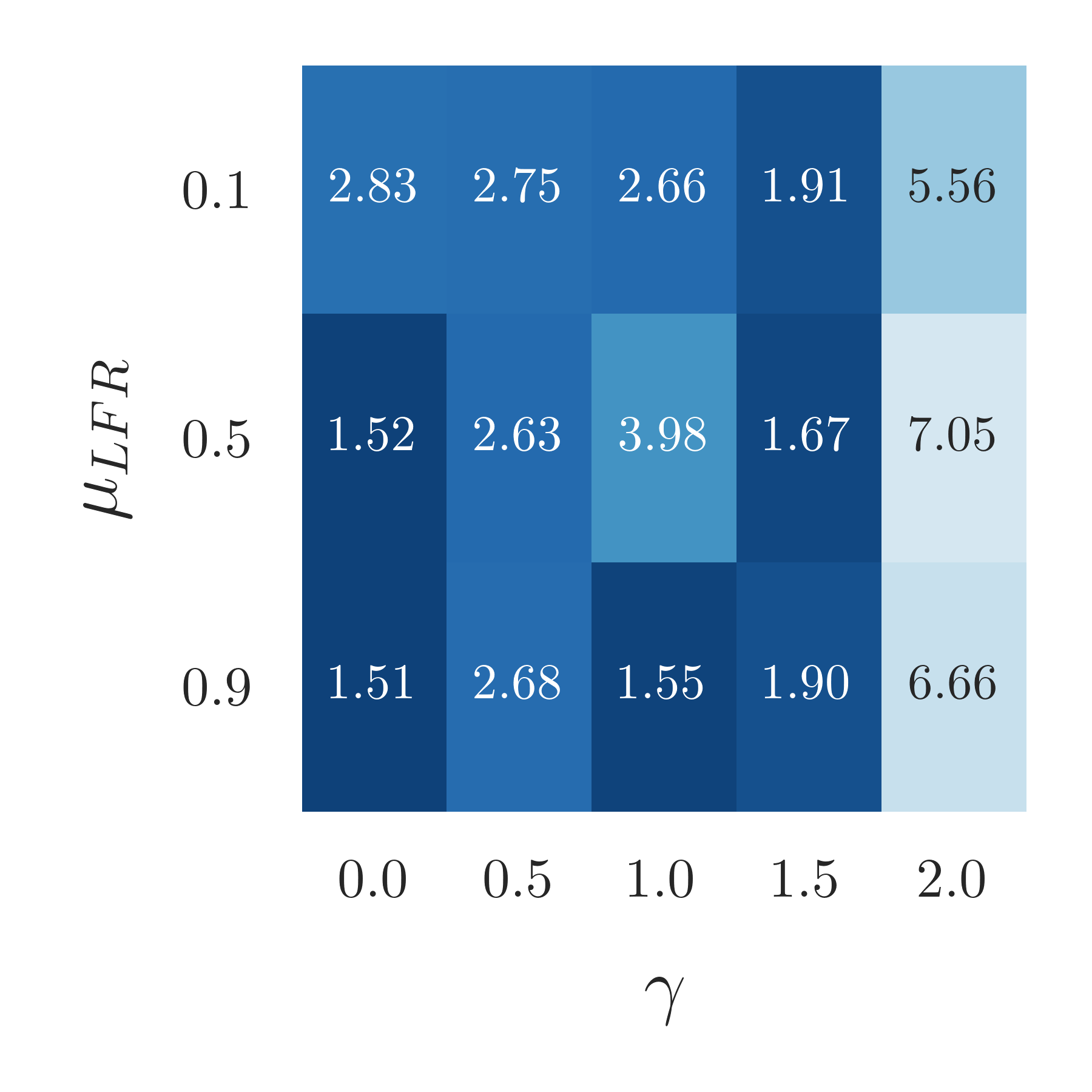}}  
\hfill
\subfloat[\centering Polarized means $\epsilon=0.2$]{ \includegraphics[width=0.3\textwidth]{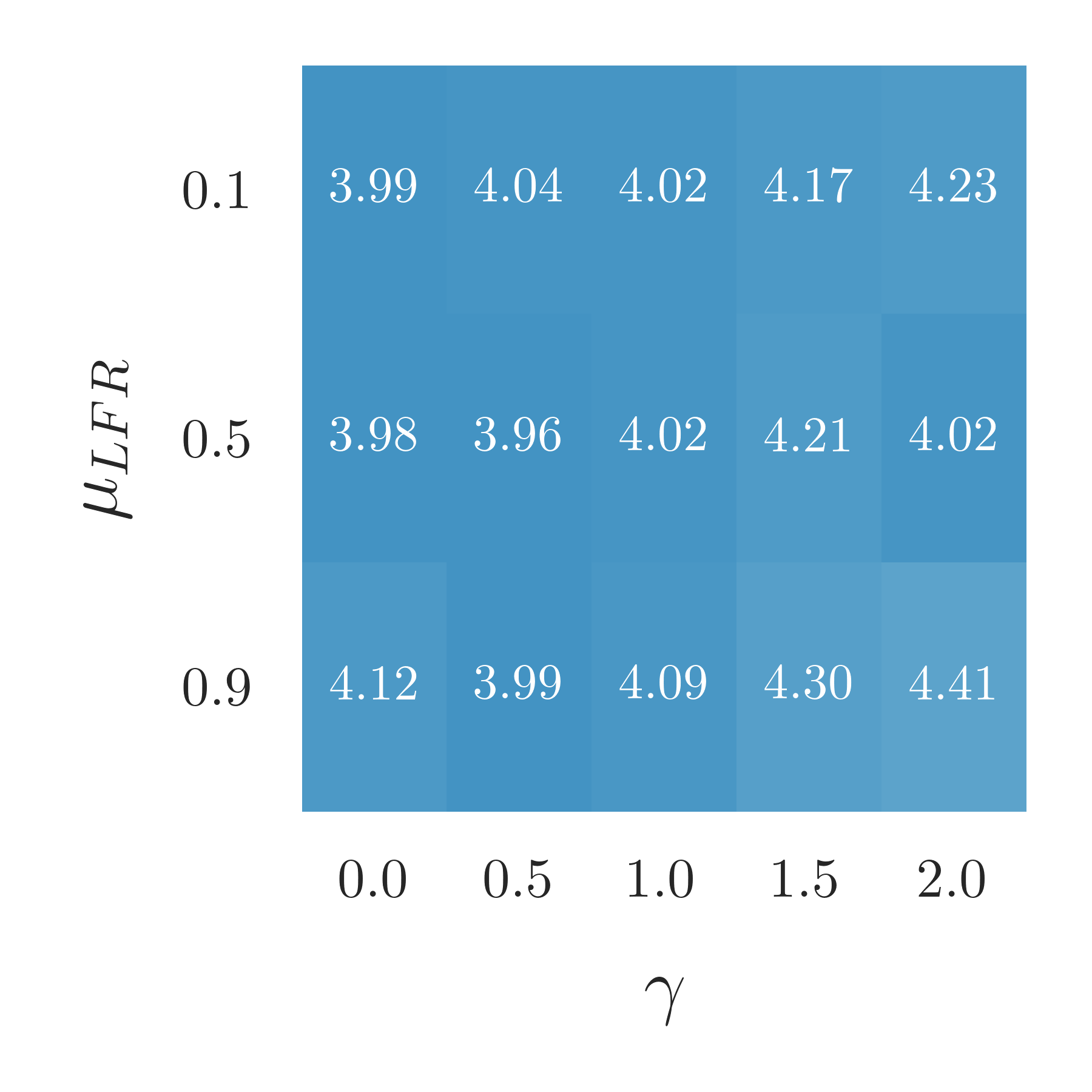}}\\

\subfloat[\centering Uniform distribution $\epsilon=0.3$]{ \includegraphics[width=0.3\textwidth]{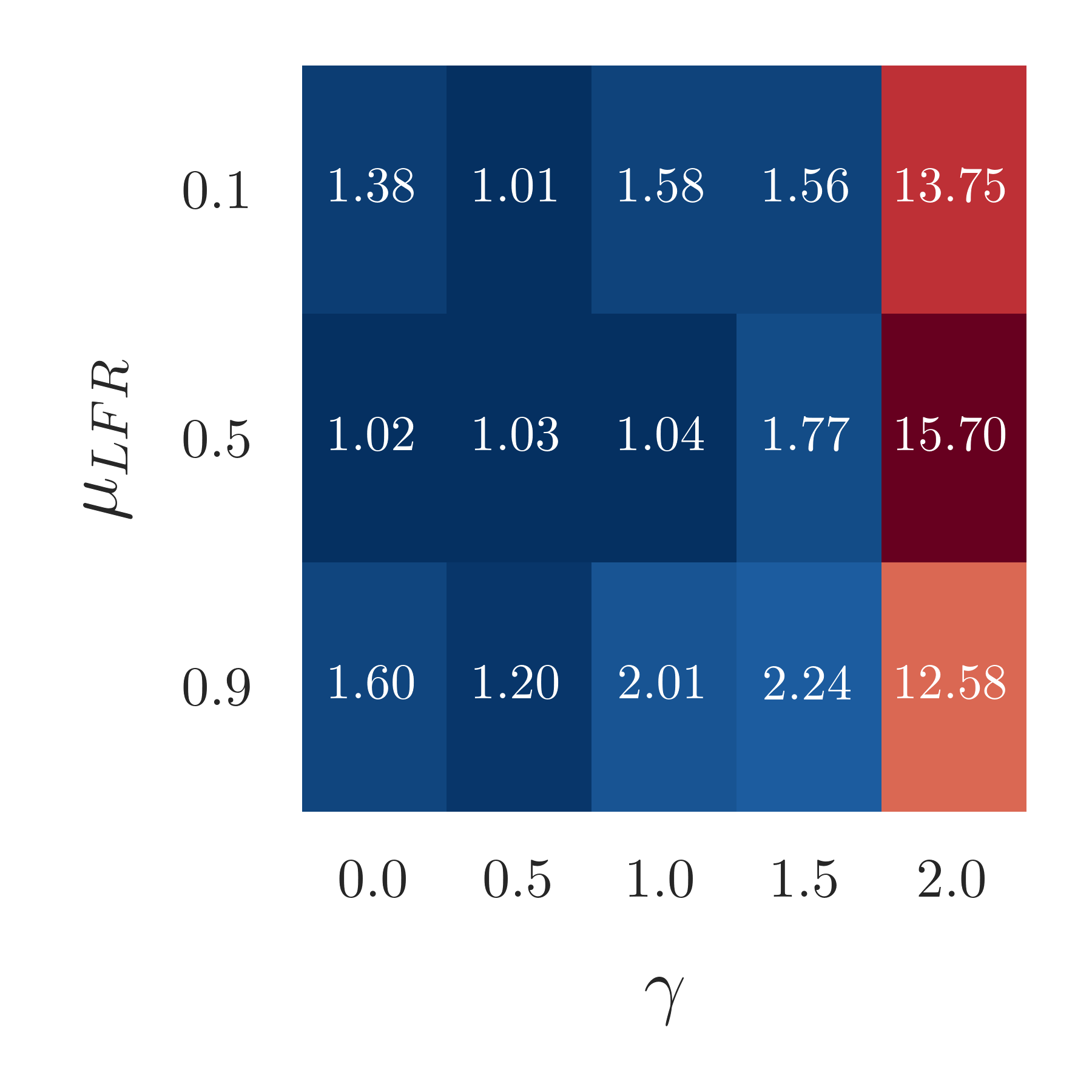}}
\hfill
\subfloat[\centering Random means {\color{white}a }$\epsilon=0.3$]{ \includegraphics[width=0.3\textwidth]{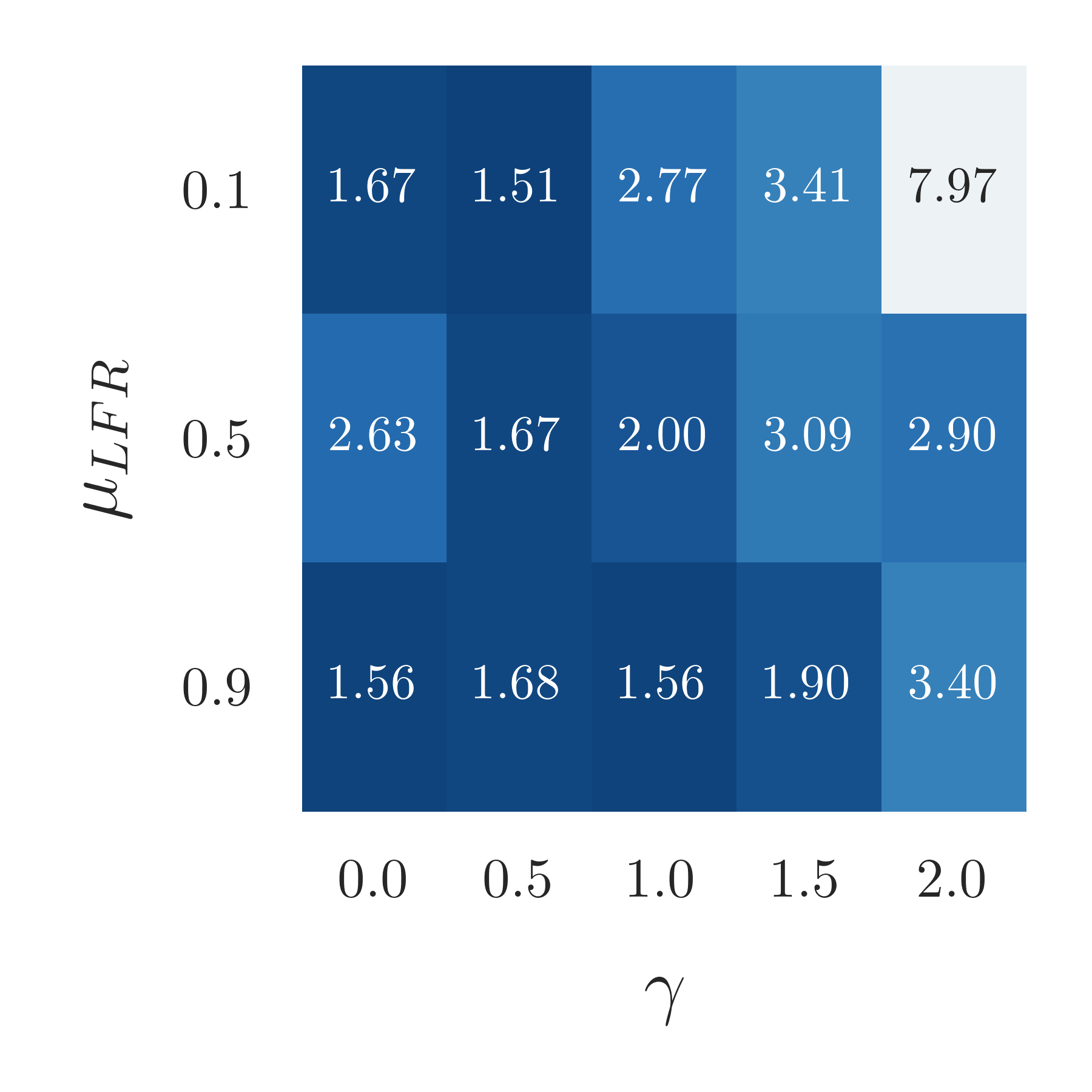}}
\hfill
\subfloat[\centering Polarized means $\epsilon=0.3$]{ \includegraphics[width=0.3\textwidth]{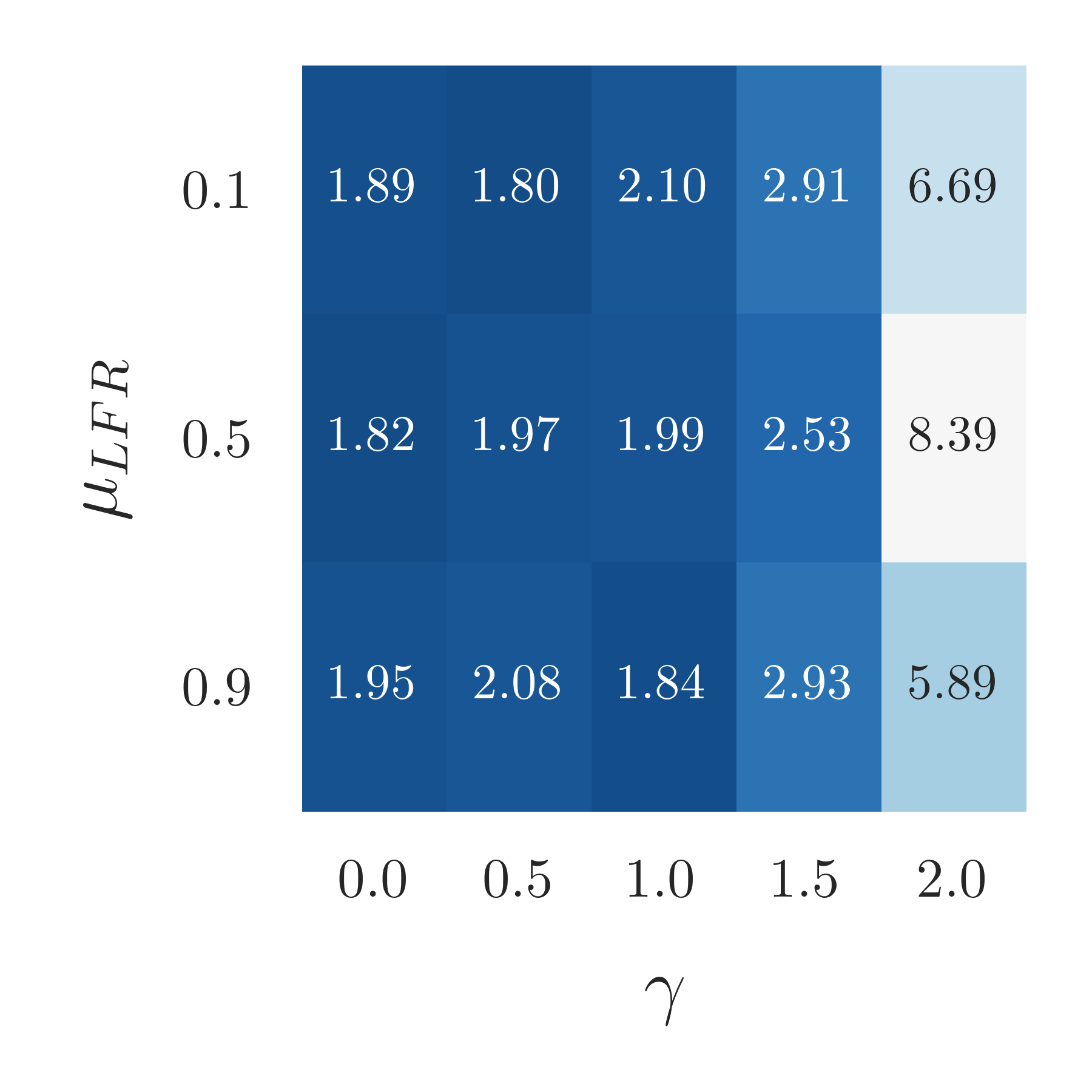}}
\hfill
\caption{\textbf{Average number of clusters for a given value of $\epsilon$ as a function of $\mu_{LFR}$ and $\gamma$.}}
    \label{fig:lfrnclusterheatmaps}
\end{figure}

From fig.\ref{fig:lfrnclusterheatmaps} it seems that the dynamics remains qualitatively the same as in the previous cases. A higher $\epsilon$ foster consensus while as $\gamma$ grows so does fragmentation. 
However, even with opinion randomly distributed across population, it seems that the mesoscale structure reduces the fragmenting effects of the bias (fig. \ref{fig:lfrnclusterheatmaps}(a)-(b)) resulting in a lower number of clusters for very high values of bias. If we start assigning to each community a random mean opinion and distributing opinions across community members with a small variance (fig. \ref{fig:lfrnclusterheatmaps}(c)-(d)) we can see that fragmentation is overall reduced. 
When the mean opinions of the communities are more distant than the confidence bound  (fig.\ref{fig:lfrnclusterheatmaps}(e)-(f)) we always obtain four to five final clusters, since different communities cannot merge and eventually, when the selection bias is very strong, some of them split into more than one cluster. However, in the case of polarized communities, if the mean initial opinions are less distant than the confidence bound, the dynamic remains is the same: we see a slow rise from polarization to fragmentation as the bias grows. 

\section{Conclusion}
\label{sec:conlcusion}
In this work, the Algorithmic Bias model - developed within the framework of bounded confidence - was simulated on complete, random scale-free network topologies. 
Such an analysis was carried out to discover and characterize the differences that affect model simulation outcomes while moving from a mean-field scenario (as proposed by the original authors) toward a more complex ones. 

Algorithmic bias is argued to be an existing factor affecting several (online) social environments. Since interactions occurring among agents embedded in such realities are far from being easily approximated by a mean-field scenario, in our study, we aimed to understand the role played by alternative network topologies on the outcome of biased opinion dynamic simulations.

From our study emerges that the qualitative dynamic of opinions remains substantially in line with what was observed assuming a mean-field context: an increase in the confidence bound $\epsilon$ favors consensus. In contrast, the introduction of the algorithmic bias $\gamma$ hinders it and favors fragmentation. 
Conversely, both simulations' time to convergence and opinions fragmentation appears to increase as the topology becomes sparser and the hub emerges.
Therefore, our analysis underlines that, alongside the algorithmic bias, the network's density heavily affects the degree of consensus reachable, assuming a population of agents with the same initial opinion distribution.

We also investigated how an underlying community structure affects the dynamics. 
What emerged is that the community structure enhances the consensus, and a larger algorithmic bias has the only effect of slowing down the convergence process. 
As already stated by the authors in \cite{SPGK19}, the initial condition is crucial to determine the final state. 
Our work showed that polarized communities that are further than the confidence bound cannot converge and that an increasing bias may favor splits into two or more clusters within the same community, even when the starting opinions were very close.
As future work, we plan to extend the AB model to cope with more realistic scenarios involving dynamic network topologies and the existence of exogenous factors affecting agents' behaviors.

\subsection*{Acknowledgments}
This work is supported by the scheme 'INFRAIA-01-2018-2019: Research and Innovation action', Grant Agreement n. 871042 'SoBigData++: European Integrated Infrastructure for Social Mining and Big Data Analytics' 

\bibliographystyle{plain}
\bibliography{references.bib}

\end{document}